\documentclass[12pt,preprint]{aastex}
\usepackage{multirow}
\usepackage{subfigure}
\usepackage{lscape}

\def\xmm{{\it XMM-Newton}}

\shortauthors{Lin et al.}
\begin{document}

\title{{\it XMM-Newton} Observations of the TeV $\gamma$-ray Source HESS J1804-216}

\author{Dacheng Lin\altaffilmark{1,2}, Natalie A. Webb\altaffilmark{1,2}, Didier Barret\altaffilmark{1,2}}
\altaffiltext{1}{CNRS, IRAP, 9 avenue du Colonel Roche, BP 44346, F-31028 Toulouse Cedex 4, France, email: Dacheng.Lin@irap.omp.eu}
\altaffiltext{2}{Universit\'{e} de Toulouse, UPS-OMP, IRAP, Toulouse, France}

\begin{abstract}
We have analyzed three {\it XMM-Newton} observations of the central
part of the unidentified TeV $\gamma$-ray source \object{HESS
  J1804-216}. We focus on two X-ray sources \object{2XMMi
  J180442.0-214221} (Src 1) and \object{2XMMi J180432.5-214009} (Src
2), which were suggested to be the possible X-ray counterparts to the
TeV source. We discover a 2.93~hr X-ray periodicity from Src 1, with
the pulse profile explained with a self-eclipsing pole in an
eclipsing polar. Src 2 exhibits a strong Fe emission line (FWHM
$\sim$0.3 keV and equivalent width $\sim$0.8 keV) and large X-ray
variability on timescales of hours and is probably an intermediate
polar. Thus Src 1 and Src 2 are probably two field sources not
responsible for the TeV emission. The observations were contaminated
by strong straylight from a nearby bright source, and we see no clear
extended X-ray emission that can be attributed to the supernova
remnant \object{G8.7-0.1}, a popular possible association with the TeV
source. The other possible association, the pulsar wind nebula
candidate \object{PSR J1803-2137}, shows little long-term variability,
compared with a previous {\it Chandra} observation. Many point sources
were serendipitously detected, but most of them are probably normal
stars. Three new candidate compact object systems (other than Src 1,
Src 2 and \object{PSR J1803-2137}) are also found. They are far away
from the TeV source and are probably also magnetic cataclysmic
variables, thus unlikely to be responsible for the TeV emission.

\end{abstract}

\keywords{acceleration of particles --- gamma rays: ISM --- X-rays: individual (2XMMi J180442.0-214221, 2XMMi J180432.5-214009, PSR J1803-2137, HESS J1804-216, G8.7-0.1)}

\section{INTRODUCTION}
\label{sec:intro}

More than 100 TeV $\gamma$-ray sources have been
discovered\footnote{http://tevcat.uchicago.edu/} \citep[See][for a
  recent review]{hiho2009}. About 60 of them, all probably Galactic,
have no firm associations at other wavelengths. \object{HESS
  J1804-216} is the brightest among these unidentified TeV
$\gamma$-ray sources, with a flux above 200 GeV around
5.3$\times$10$^{-11}$ erg~s$^{-1}$~cm$^{-2}$ (nearly 25\% of the Crab
Nebula). It was discovered by the High Energy Stereoscopic System
(H.E.S.S.) in 2004, with the best-fit position of
$l$=$8.401\degr\pm0.016\degr$ and $b$=$-0.033\degr\pm0.018\degr$ in
Galactic coordinates \citep{ahakay2005,ahakba2006}. It is extended,
with a radius of $\sim$12$\arcmin$.

The Galactic TeV sources with firm associations include high-mass
X-ray binaries (HMXBs), supernova remnants (SNRs), and pulsar wind
nebulae (PWNe), and they can be point-like (mostly HMXBs) or extended
\citep[mostly SNRs and PWNe,][]{kapaga2007,hiho2009}. Extragalactic
TeV sources are currently represented only by active galactic nuclei
(AGN) and appear point-like, except the starburst galaxies, M 82 and
NGC 253, whose TeV emission was also observed but could not be
resolved by current instruments \citep{acahak2009,
  veacal2009}. \object{HESS J1804-216} is most likely a Galactic
source, considering its extended nature and low Galactic latitude. It
has several possible associations suggested, though none of them are
very compelling. One candidate is the large (with a size of
$\sim45\arcmin$ in radio) and low-surface-brightness SNR
\object{G8.7-0.1} \citep{kawe1990}, as \object{HESS J1804-216} has
some overlap with the southwestern part of its shell
\citep{ahakba2006}. Another candidate is the young Vela-like pulsar
\object{PSR J1803-2137} \citep[the spin period is 134
  ms,][]{clly1986}, which is $\sim$11$\arcmin$ away from the best-fit
position of \object{HESS J1804-216}. \citet{ajalba2012} reported two
GeV sources (their sources E and W) detected with the {\it Fermi}
Large Area Telescope (LAT) around \object{HESS J1804-216} but found
that they do not match with the TeV source in morphology or
spectrum. Instead, their source E has most emission positionally
coincident with \object{SNR G8.7-0.1}.

Although both \object{SNR G8.7-0.1} and \object{PSR J1803-2137} are
also X-ray sources beside their radio emission, there are other X-ray
sources that are positionally closer to \object{HESS J1804-216} than
them. \citet{bakohi2007} reported detection of two highly absorbed
hard X-ray sources with {\it Suzaku}, i.e., \object{Suzaku J1804-2142}
(our Src 1 below) and \object{Suzaku J1804-2140} (our Src 2 below),
only $\sim$2$\arcmin$ away from the best-fit position of \object{HESS
  J1804-216} and suggested them as the plausible X-ray counterparts to
this TeV source. \citet{kapaga2007} analyzed the {\it Chandra}
observation of these two X-ray sources (\object{CXOU J180441.9-214224}
and \object{CXOU J180432.4-214009}, respectively) and obtained similar
results.

In \citet[][LWB12 hereafter]{liweba2012}, we carried out source type
classifications of 4330 sources from the 2XMMi-DR3 catalog
\citep{wascfy2009}, which is based on \xmm\ pointed observations. In
that project, we serendipitously discovered a 2.93~hr X-ray
periodicity from Src 1 and a strong but narrow Fe emission line from
Src 2 from three \xmm\ observations of the central part of
\object{HESS J1804-216}. In this work, after excluding Src 1 and Src 2
as the X-ray counterparts to \object{HESS J1804-216}, we search for
other possible counterparts in the field of view (FOV), using the
source type classification scheme from LWB12.

In Section~\ref{sec:reduction}, we describe the screening of spurious
sources along with detailed spectral and timing analysis of the
relatively bright sources that we selected for final study. In
Section~\ref{sec:res}, we present the results of the source
classification and the detailed properties of the sources. In
Section~\ref{sec:dis}, we discuss the possible nature of the sources
and the possible X-ray counterparts to \object{HESS J1804-216}. The
conclusions of our study are given in Section~\ref{sec:conclusion}.

\section{DATA ANALYSIS}
\label{sec:reduction}

\subsection{Source Detection and Screening}

\begin{figure}[]
\centering
\includegraphics[width=6.8in]{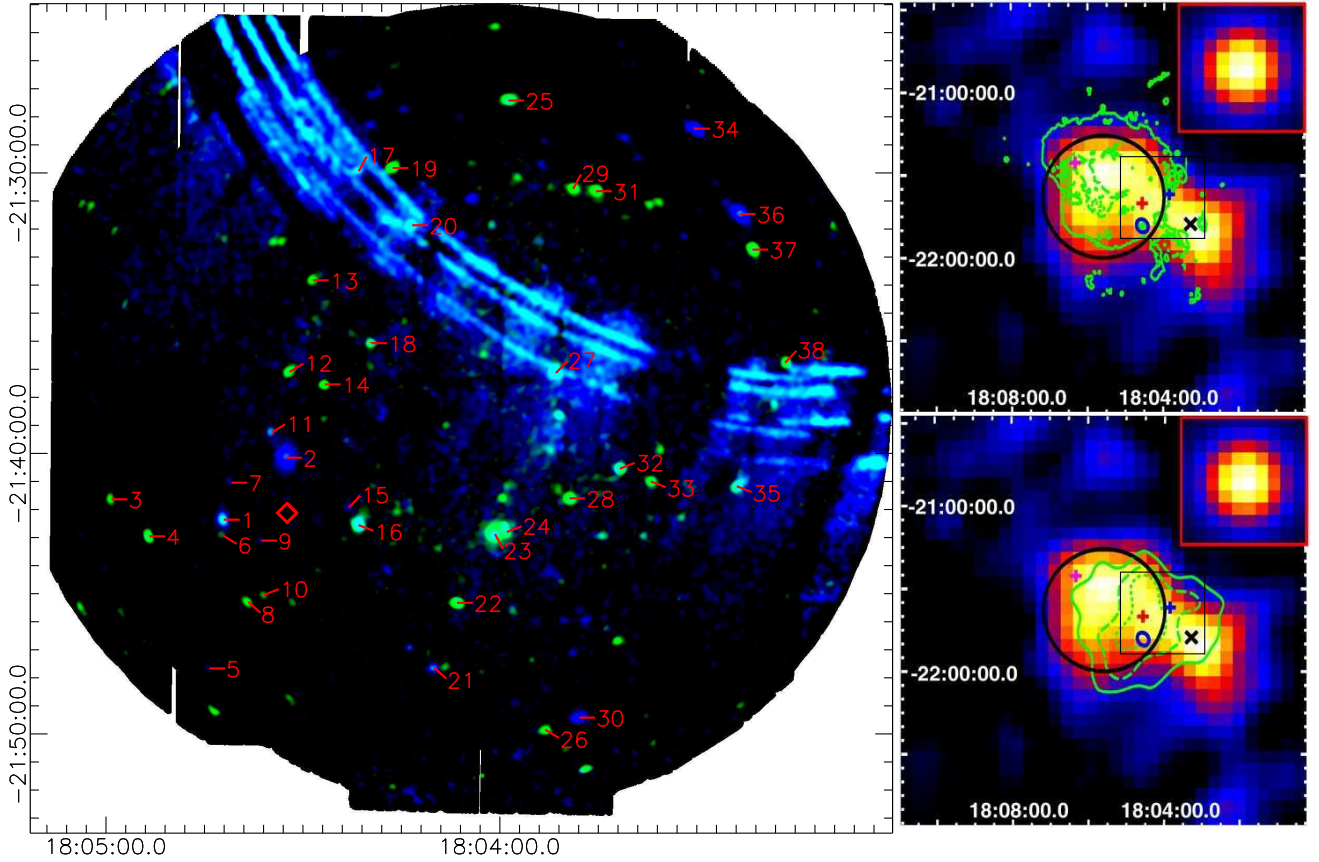}
\caption{(Left panel): The \xmm\ X-ray image of the field around HESS~J1804-216. Data from all three EPIC cameras from all three observations were combined. The image is false-colored as follows: 0.5--2.0 keV photons in green and 2.0--12.0 keV photons in blue (We did not include photons at lower energies to avoid low-energy noise). The image has a bin size of 2$\farcs$2 and is smoothed by a Gaussian kernel of $\sigma$=5$\arcsec$. The red diamond marks the best-fit position of HESS~J1804-216. The sources studied by us are numbered from the left to the right, except Src 1 and Src 2; see Table~\ref{tbl:dercat}. (Right panels): The {\it Fermi} LAT 2--10 GeV map from \citet{ajalba2012}, with the green contours indicating the VLA 90 cm image at 5\%, 15\%, and 25\% of the peak intensity \citep{brgega2006} in the top panel and the subtracted TeV photon counts of HESS J1804-216 at 25\%, 50\%, and 75\% levels \citep{ahakba2006} in the bottom panel. The black box corresponds to the size of the left panel. The black circle indicates the best-fit disk size of Source E in \citet{ahakba2006}, while the black cross is for their Source W. The blue, magenta, and red plus signs indicate PSR J1803-2137 (our Src 27), PSR J1806-2125, and Suzaku J1804-2140 (our Src 2), respectively, while the blue ellipse indicates the radio extension of SNR G8.31-0.09 \citep{ahakba2006}.  \label{fig:colorimage}}
\end{figure}

The three \xmm\ observations that covered the central part of
\object{HESS J1804-216} are 0503170101, 0503170201 and
0503170301. They will be referred to as Obs 1, 2, and 3,
respectively. Obs 1 was made on 2007 October 2, lasting for about 54
ks, while Obs 2--3 were made only six days later, i.e., 2007 October
8--9, lasting for about 31 ks each. These three observations had the
same pointing direction, and the three European Photon Imaging Cameras
(EPIC), i.e., pn, MOS1/M1, and MOS2/M2
\citep{jalual2001,stbrde2001,tuabar2001}, all used the ``medium''
optical blocking filter and the ``full frame'' imaging mode. The X-ray
image combined from all these observations and the corresponding FOV
relative to observations in other wavelengths are shown in
Figure~\ref{fig:colorimage}.

Sources detected from these three observations are included in the
2XMMi-DR3 catalog, with 524 in total. We found from visual inspection
that the catalog has successfully included all significant
sources. However, there are also many spurious sources because they
appear in the straylight structures (see the bright arcs and the blue
stripes in the EPIC image in Figure~\ref{fig:colorimage}) caused by
single reflections of a bright X-ray source outside the FOV. There are
also other spurious sources due to problems of low-energy noise near
the central line in the pn camera (such sources are detected only in
the pn image and only below 0.5 keV) and noisy CCDs in both MOS
cameras (such sources tend to be detected as extended in the 2XMMi-DR3
catalog) in all three observations. Because the straylight structures
tend not to completely match in sky position in different cameras and
noisy CCDs are camera specific, the spurious sources caused by these
problems can show no emission in cameras that are free of these
problems at the position of the sources, which can be used to confirm
their spurious nature. We find from visual screening that about 150
sources are probably real. However, most of them are faint and have
low statistics, making it difficult to study their source
properties. Thus in this study we only concentrate on 38 sources that
have the signal-to-noise ratio S$/$N $>8$, or S$/$N $>5$ if they are
within 3$\arcmin$ from the best-fit position of \object{HESS
  J1804-216}, in at least one observation
(Table~\ref{tbl:dercat}). Our numbering of these sources is given in
Figure~\ref{fig:colorimage} and Table~\ref{tbl:dercat}.

We followed the procedure in LWB12 to search for the optical and IR
counterparts to these 38 sources in the USNO-B1.0 Catalog
\citep{moleca2003} and the 2MASS Point Source Catalog \citep[2MASS
  PSC,][]{cuskva2003} and calculated their X-ray-to-optical and
X-ray-to-IR flux ratios, respectively (Table~\ref{tbl:dercat}). The
counterpart is chosen to be the closest one within 2$\arcsec$ or three
times the X-ray positional error, whichever is larger. We also
followed the LWB12 procedure to fit the spectra created from the band
count rates included in the 2XMMi-DR3 catalog with an absorbed
powerlaw (PL), in order to roughly characterize the spectral shape of
the sources.

\subsection{Detailed Spectral and Timing Analysis}
We created light curves for each source and visually searched for
stellar X-ray flares to help the source classification. We used SAS
11.0.0 and the calibration files of 2011 June to reprocess the X-ray
event files and do the follow-up analysis. Very short intervals of
strong background flares occured in the pn camera in all three
observations and were excluded following the SAS thread for filtering
high background. The final pn/MOS1/MOS2 exposures used are 46/54/54,
22/31/31, and 25/32/32 ks for Obs 1-3, respectively. For all sources,
we used a circular source region centered on the source (the radii
used for Src 1 and Src 2 are 15$\arcsec$ and 20$\arcsec$,
respectively; for others, see Figure~\ref{fig:lcsum}). The background
was estimated from a large circular region, typically with a radius of
50$\arcsec$--100$\arcsec$, near the source in each camera. The event
selection criteria followed the default values in the pipeline (see
Table~5 in \citet{wascfy2009}).

We carried out detailed spectral fitting and timing analysis for
sources that we find to be possibly compact object systems containing
a white dwarf (WD), a neutron star (NS), or a stellar-mass black hole
(BH). As their spectra are in general hard, we only used an absorbed
PL to fit their continuum. To improve the statistics and considering
that Obs 2 and Obs 3 are relatively short and close in time (in one
day), the spectra of both observations were added together for each
camera. We rebinned the spectra to have at least 20 counts in each bin
so as to adopt the $\chi^2$ statistic for the spectral fits.

To search for X-ray periodicity, we created the Leahy power
\citep{ledael1983} from the light curves that combined all EPIC
cameras and were extracted from the source region. We used a binsize
of 2.6 s (the frame readout time of the MOS cameras). Although the pn
camera has a higher time resolution (73.4 ms) and using data
exclusively from this camera allows to search for periodicities at
higher frequencies, we found no powers above a 99.9\% confidence level
at high frequencies. Thus only results using light curves binned at
2.6 s and combining all EPIC cameras will be shown.

To constrain the values of their periods and examine the profiles of
the modulation, we employed the epoch folding search technique. The
phase folded background-subtracted light curve was obtained by
subtracting the folded light curve from the background region (after
being rescaled to the size of the source region) from the folded light
curve from the source region combining all EPIC cameras.

\section{RESULTS}
\label{sec:res}

\subsection{X-ray Properties of Src 1}
The spectra of Src 1 are well fitted with an absorbed PL model with
photon indices $\Gamma_{\rm PL}\sim 1.0$ and the column density
$N_{\rm H}\sim 10^{22}$ cm$^{-2}$ (Table~\ref{tbl:spfit}). The average
flux in Obs 1 is slightly higher (a few tens of percent) than that in
Obs 2--3.

\begin{figure*} 
\centering
\includegraphics[width=6.8in]{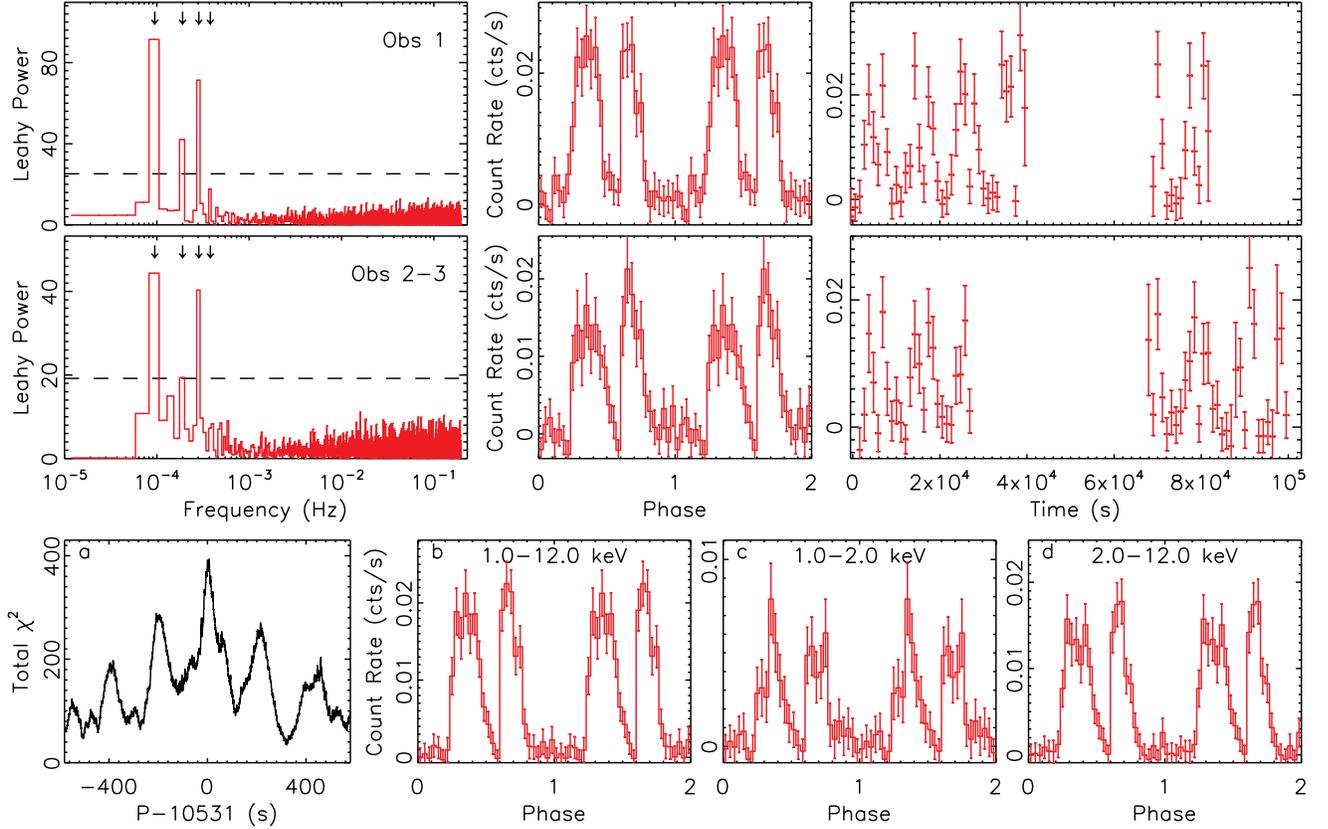}
\caption{The timing properties of Src 1. The panels in the top two rows show the Leahy power (left), the light curve folded at a period of $P_0$=10531 s (middle), and the (unfolded) light curve (right) for each observation (1.0--12.0 keV). The arrows from the left to the right in the power plots mark the period $P_0$ and the harmonics $P_0/2$, $P_0/3$, and $P_0/4$, respectively. The dashed lines indicate a 99.9\% confidence detection level. The unfolded light curves are shifted in time to be aligned in phase. Panel (a) in the bottom row shows the total $\chi^2$ values from the fits to a constant to the 1.0--12.0 keV light curves folded at various tentative periods using all three observations. The 1.0--12.0 keV, 1.0--2.0 keV and 2.0--12.0 keV light curves folded at $P_0$ also using all three observations are given in panels (b), (c), and (d), respectively. \label{fig:254026foldcurve}}
\end{figure*}

The timing properties of Src 1 are shown in
Figure~\ref{fig:254026foldcurve}, and we see strong powers at a period
of $P_0$$\approx$10531~s (i.e., 2.93~hr) and its harmonics (especially
$P_0$/3) in both Obs 1 and Obs 2--3. Such a periodic behavior can also
be seen in the light curves in the same figure. The pulse profile
shows double peaks with comparable maxima and does not seem to vary
over the observations (Figure~\ref{fig:254026foldcurve}). The count
rate is consistent with zero in the two low phases, with one lasting
for about one third of the period and the other very brief ($\sim$3\%
of the period). The pulse profile seems to vary slightly with energy.

\subsection{X-ray Properties of Src 2}
The fits to the spectra of Src 2 with an absorbed PL show strong
residuals around the Fe line energy 6--7 keV
(Figure~\ref{fig:s253961spfits}), resulting in high values of
$\chi_\nu^2$ ($\sim$1.5, Table~\ref{tbl:spfit}). Thus we added a
Gaussian line to the fits. We saw no significant variation of the line
centroid energy $E_{\rm Ga}$ and FWHM $\sigma_{\rm Ga}$ between Obs 1
and Obs 2--3. Thus we chose to tie these parameters to be the same in
these observations in the final fits, to have better constraints on
them. We obtained good fits with small photon indices ($\sim$0.7) and
large equivalent widths (EWs) of the Fe line ($\sim$0.8 keV) in both
Obs 1 and Obs 2--3 (Table~\ref{tbl:spfit} and
Figure~\ref{fig:s253961spfits}).

The timing properties of Src 2 are shown in
Figure~\ref{fig:253961foldcurve}. We see large variability of the
source on timescales of about ten hours. The epoch folding search
technique suggests a $\sim$11.4 hr period
(Figure~\ref{fig:253961foldcurve}), which, however, needs to be
confirmed with longer observations.

\subsection{Properties of Other Sources}
The results of multi-wavelength cross-correlation and simple X-ray
spectral fits are given in Table~\ref{tbl:dercat}. Optical and IR
counterparts were found for most of the 38 sources. The light curves
of 14 sources with some flaring activity are shown in the bottom
panels of Figure~\ref{fig:lcsum}. Stellar flares tend to show a fast
rise and a slow decay, as we see in sources like Src 7, 18, and 25,
while most of the others do not have such a structure but only exhibit
the rise or the decay, due to observation gaps. In LWB12, sources with
the X-ray-to-IR flux ratio logarithm $\log(F_{\rm X}/F_{\rm IR})$
(Table~\ref{tbl:dercat}) less than $-0.9$ and/or with stellar X-ray
flares are classified as stars. Following this scheme, we have 32
candidate stars, based on Table~\ref{tbl:dercat} and
Figure~\ref{fig:lcsum}. These stars generally have very soft X-ray
spectra ($\Gamma_{\rm PL}\gtrsim 4$ in Table~\ref{tbl:dercat}) and
thus mostly appear green in the X-ray colored image in
Figure~\ref{fig:colorimage}. Their spectra become harder when there
are stellar flares in the light curves. Some sources (e.g., Src 4--7)
were detected because of flares.

The other six sources (Src 1, 2, 27, 30, 34 and 36) are candidate
compact objects. We ruled them out as being AGN, considering that they
appear in the Galactic plane and that their X-ray spectra are hard
(LWB12). Src 27 is \object{PSR J1803-2137}. Similar to Src 1 and Src
2, we calculated the Leahy powers for the latter four sources but
found no strong power above the 99.9\%-confidence level in the
frequency range of $10^{-5}$--0.5 Hz. The light curves, plotted in
Figure~\ref{fig:lcsum}, exhibit little variability. We note that for
\object{PSR J1803-2137} we did not use the pn data because the source
falls across the CCD gap and near the bright arc
(Figure~\ref{fig:colorimage}) in all observations.

We also carried out detailed spectral fits for these sources. The
results are given in Table~\ref{tbl:spfit}. Src 30, 34 and 36 are all
hard ($\Gamma_{\rm PL}\lesssim 1$) and heavily absorbed ($N_{\rm
  H}\sim 5\times$$10^{22}$ cm$^{-2}$). For Src 34 and Src 36, adding a
narrow Fe line improves the fits, similar to Src 2.

In Figure~\ref{fig:colorimage}, many other (fainter) point sources can
be seen. Most of them appear green (thus soft) and are probably
stars. We note that the 2XMMi-DR3 catalog also includes 113 extended
sources (with the extent radius greater than 6$\arcsec$ in at least
one of the three observations). However, they are all probably
spurious, because most of them appear in the straylight structures,
while others (typically with S$/$N around 5) seem to be caused by
noisy MOS CCDs or bright sources. Thus none of these extended sources
are included in our 38 sources.

\section{DISCUSSION}
\label{sec:dis}

\subsection{Comparison with Previous X-ray Study of Src 1 and Src 2}
There are studies of observations of the field around \object{HESS
  J1804-216} by other X-ray observatories. Here we briefly compare
their results with ours, focusing on the properties of Src 1 and Src
2. In the {\it Chandra} 2005-05-04 observation, \citet{kapaga2007} saw
a weak and possibly extended/multiple source (\object{CXOU
  J180441.9-214224}) with a size of 1.5$\arcmin$--2$\arcmin$ and the
centriod only 3$\arcsec$ away from Src 1. They also found the source
\object{CXOU J180432.4-214009}, which has a 95\% positional error of
1$\farcs$4 and is only 0\farcs9 away from Src 2, based on the position
information from the {\it Chandra} Source Catalog \citep[release
  1.1,][]{evprgl2010}. In the {\it Swift} 2005-11-03 observation (12 ks)
studied by \citet{labama2006}, no source near Src 1 was detected above
3$\sigma$, but they found a source (their Src 3 (Sw3 hereafter), with
a positional error of 5$\arcsec$) 7$\arcsec$ away from Src 2. In the
{\it Suzaku} 2006-04-06 observation (40 ks), \citet{bakohi2007} found
two sources, \object{Suzaku J1804-2142} and \object{Suzaku
  J1804-2140}, 34$\arcsec$ and 23$\arcsec$ away from Src 1 and Src 2,
respectively (the {\it Suzaku} systematic positional uncertainty is
1$\arcmin$), with \object{Suzaku J1804-2140} claimed to be
extended/multiple. Based on the positional coincidence, \object{CXOU
  J180441.9-214224} and \object{Suzaku J1804-2142} are probably the
same source as Src 1, and \object{CXOU J180432.4-214009}, Sw3, and
\object{Suzaku J1804-2140} are probably the same source as Src
2. Considering that both Src 1 and Src 2 are point-like with the
extent radius $<6\arcsec$ from the 2XMMi-DR3 catalog, which is
supported by the large X-ray modulations on timescales of hours seen
in both sources, the extended/multiple feature of \object{CXOU
  J180441.9-214224} and \object{Suzaku J1804-2140} are probably due to
contamination emission from nearby stars or noisy background. For
\object{CXOU J180441.9-214224}, the contaminating source could be the
flaring star Src 6 (34$\arcsec$ away from Src 1), and for
\object{Suzaku J1804-2140}, it could be the bright star Src 11
($\sim1\arcmin$ away from Src 2), which seems to be persistent and has
no flare observed by us.

The fluxes of the candidate counterparts to Src 1 and Src 2 estimated
by the above studies are within a factor of about two of those that we
obtained here, indicating a fairly constant flux level for both
sources over about two years. This is not in conflict with the
non-detection of Src 1 in the {\it Swift} 2005-11-03 observation,
considering that this observation only detected Src 2, which has a
flux five times more than Src 1, at only 4$\sigma$. \citet{bakohi2007}
fitted the spectrum of \object{Suzaku J1804-2142} with an absorbed PL
and inferred a harder spectrum ($\Gamma_{\rm PL}=-0.3\pm0.5$) than Src
1, but this could be caused by background contamination, which could
be a serious issue due to the low statistics of \object{Suzaku
  J1804-2142}. The fit of \object{CXOU J180432.4-214009} by
\citet{kapaga2007} and that of \object{Suzaku J1804-2140} by
\citet{bakohi2007} using an absorbed PL both inferred an absorption
column density and a photon index with large error bars but fully
consistent with those that we obtained for Src 2. We note that
\citet{kapaga2007} and \citet{bakohi2007} did not need an Fe line in
their spectral fits, probably due to the low statistics of their
data. Concerning the timing properties, \citet{kapaga2007} found a 106
s period candidate at only 2.3$\sigma$ for \object{CXOU
  J180432.4-214009}, which we were unable to confirm in either Obs 1
or Obs 2--3 of Src 2 (Figure~\ref{fig:253961foldcurve}). It could be
because this periodicity is not real or is intermittent.

\subsection{Nature of Src 1, 2, 30, 34 and 36}
Considering that the Galactic TeV sources with firm associations are
mostly HMXBs, SNRs and PWNe, \citet{bakohi2007} and \citet{kapaga2007}
discussed both Src 1 and Src 2 in terms of these source types as the
X-ray counterparts to \object{HESS J1804-216}. However, the large
X-ray modulation on timescales of hours observed by us in these two
sources excludes them from being SNRs or PWNe.

It is unlikely that Src 1 is a HMXB for several reasons. Its 2.93~hr
X-ray period is too short to be the orbital period, because HMXBs
mostly have orbital periods longer than one day
\citep{bichch1997,chco2006} (but this cannot exclude it as a low-mass
X-ray binary like the pulsar \object{Her X-1}). This period seems too
long to be the NS spin period either (the slowest pulsar known is
\object{RX J0146.9+6121}, with a spin period of 1412~s
\citep{he1994}). It has a long low phase in the pulse profile, which
is rarely seen in HMXBs \citep[e.g.,][]{bichch1997}. Its 0.3--10.0 keV
unabsorbed luminosity is about $10^{33}$ erg s$^{-1}$ (assuming a
source distance of 8 kpc), which is too low for accretion-powered
X-ray pulsars \citep[typically higher than $10^{34}$ erg
  s$^{-1}$,][]{muarba2004}. Considering the low luminosity and the
hard spectra, we suggest Src 1 to be a magnetic cataclysmic variable
(CV; polar or intermediate polar). The 2.93~hr period could be either
the orbital period or the spin period of the WD (they are the same for
polars). Further, the observed pulse profile is similar to those of
some eclipsing polars at high energies (above 1 keV), e.g., the
eclipsing polar \object{HU Aquarii} \citep[See Figure~3
  in][]{scscvo2009}. The light curve would then be dominated by a
single bright phase interval (phase 0.2--0.8 in
Figure~\ref{fig:254026foldcurve}) of a self-eclipsing pole, with the
short low period (phase around 0.6) caused by the eclipse of the
accretion region by the secondary star. We checked whether an Fe
emission line is present in the spectra of Src 1 by adding a Gaussian
emission line to the spectral fits, considering that magnetic CVs
often exhibit Fe emission lines with EWs typical of a few hundred eV
\citep{ezis1999}. We found that adding the line did not improve the
fits (the total $\chi^2$ values decreased by less than 2 for both Obs
1 and Obs 2-3), but we also obtained very high EW upper limits
($\sim$1.4 keV, at a 90\% confidence level). Thus we cannot rule out
the presence of Fe emission lines in Src 1 at the level often seen in
magnetic CVs.

Src 2 has a 0.3--10.0 keV unabsorbed luminosity of about
6$\times$$10^{33}$ erg s$^{-1}$, assuming a source distance of 8
kpc. Such a luminosity also seems low for a HMXB. Its Fe emission line
is also stronger than most HMXBs \citep[EW$\lesssim$100
  eV,][]{muarba2004}, while it is close to typical values seen in
magnetic CVs \citep{ezis1999}. In fact, sources with an X-ray
luminosity, a hard spectrum, and a strong Fe emission line like Src 2
are commonly seen toward the Galactic center, and \citet{muarba2004}
argued that they are most likely intermediate polars. Src 2 is
probably one such object.

Src 30, 34 and 36 have similar low luminosities (also assuming a
distance of 8 kpc) and hard spectra like Src 1 and Src 2. Thus they
are probably magnetic CVs too. Although they have steady light curves
and may be rotation-powered pulsars like \object{PSR J1803-2137}, we
do not favor this interpretation because the rotation-powered pulsars
tend to have $\Gamma_{\rm PL}\gtrsim 1$ \citep[][LWB12]{muarba2004}.

\subsection{The X-ray Counterpart to HESS J1804-216}
If Src 1 and Src 2 are CVs, they are unlikely to be the counterparts
to \object{HESS J1804-216}. Src 30, 34 and 36 are far away (more than
12$\arcmin$) from \object{HESS J1804-216} and are unlikely to be its
counterparts either. The promising counterpart candidates are the SNR
\object{G8.7-0.1} and \object{PSR J1803-2137} (i.e., Src 27), although
they are not positionally close to this TeV source
\citep{ahakba2006}. \citet{ajalba2012} explained the GeV and TeV
$\gamma$-ray emission in terms of the interaction between
\object{G8.7-0.1} and adjacent (for GeV emission) or distant (for TeV
emission) molecular clouds \citep{blfist1982}, which is supported by
the detection of a single bright OH(1720 MHz) maser in the eastern
edge of \object{G8.7-0.1} by \citet{heyu2009}.

{\it ROSAT} detected extended thermal X-ray emission from the
northeastern part of \object{G8.7-0.1}, with a temperature of 0.3--0.7
keV and a column density of (1.2--1.4)$\times$10$^{22}$ cm$^{-2}$
\citep{fioe1994}. \citet{pufrsa2011} obtained consistent results and
supported the thermal nature of the extended X-ray emission using the
{\it XMM-Newton} observation 0405750201 (exposure 16 ks). The X-ray
image of this observation is shown in Figure~\ref{fig:colorimage_ne}
(available in the electronic version of the article), in which the
extended thermal emission appears in green. There is no clear extended
hard emission other than the straylight, which is also seen in this
observation, and the small extended source near \object{PSR
  J1806-2125} (the magenta plus in Figure~\ref{fig:colorimage_ne}),
which has an extent radius of only $21.1\pm0.7\arcsec$ from the
2XMMi-DR3 catalog. This extended emission was suggested to be a new
PWN unassociated with \object{PSR J1806-2125} by \citet{pufrsa2011}.

In spite of six times more exposure, Obs 1-3 failed to detect extended
thermal X-ray emission as seen in observation 0405750201. This could
be due to spatial variation in absorption, which, however, was not
supported by \citet{fioe1994}, because they observed no significant
spatial correlation between the intensity and the hardness ratio of
the X-ray emission. Alternatively, X-ray emission from the SNR can be
intrinsically very weak, considering its low surface brightness.

\citet{ajalba2012} could not rule out the possibility that the TeV
emission is from the inverse Compton scattering of the relativistic
electrons in a PWN associated with sources such as \object{PSR
  J1803-2137} and \object{Suzaku J1804-2140} (i.e., our Src 2). We
have argued that Src 2 is not a PWN. \object{PSR J1803-2137} is
extended but has a diameter size of only $\sim$7$\arcsec$ from the
{\it Chandra} observation \citep{cuko2006}. This size is too small to
be resolved by {\it XMM-Newton}. The spectral parameters that we
obtained from the {\it XMM-Newton} observations are fully consistent
with those obtained by \citet{cuko2006}, indicating little long-term
X-ray variability. This is expected for rotation-powered pulsars
(LWB12). The short-term variability is small too
(Figure~\ref{fig:lcsum}), as expected for an extended
source. \citet{cuko2006} suggested that \object{PSR J1803-2137} is a
PWN, whose X-ray emission probably just reflects the bright core of
the (large) PWN. We note that \object{PSR J1803-2137} was suggested to
be associated with \object{G8.7-0.1} \citep{kawe1990}, but this was
not supported by the proper motion of \object{PSR J1803-2137} measured
by \citet{brcaku2006}, which indicates a birth position at the extreme
edge of the SNR.

\section{CONCLUSIONS}
\label{sec:conclusion}
We have studied three {\it XMM-Newton} observations of the central
part of the unidentified TeV $\gamma$-ray source \object{HESS
  J1804-216}. We focus on two X-ray sources (Src 1 and Src 2) that
were suggested to be the counterparts to the TeV source. Src 1 is
probably an eclipsing polar, based on our discovery of a 2.93~hr X-ray
periodicity from this source and the pulse profile that can be
naturally explained by a self-eclipsing pole. We detect a strong Fe
emission line (FWHM $\sim$ 0.3 keV and EW $\sim$ 0.8 keV) and strong
X-ray variability on timescales of hours from Src 2. This source is
probably an intermediate polar, which is often seen in the Galactic
center. Thus Src 1 and Src 2 are probably two magnetic CVs that happen
to be in the direction of the TeV source, not its X-ray counterparts.

We carry out systematic classification of 38 relatively bright sources
and find three new compact object systems, in addition to Src 1, Src 2
and \object{PSR J1803-2137}. Others are probably normal stars. The new
compact object systems are probably CVs too, and they are far away
from \object{HESS J1804-216}, thus unlikely to be its X-ray
counterparts either. \object{HESS J1804-216} is still most likely
associated with SNR \object{G8.7-0.1} or \object{PSR J1803-2137}, as
suggested before. No clear extended emission is observed, confirming
that \object{G8.7-0.1} only has bright thermal emission in its
northeastern shell. \object{PSR J1803-2137} shows little long-term
X-ray variability compared with previous studies and cannot be
resolved by {\it XMM-Newton}, indicating an extent radius
$<$6$\arcsec$.

\acknowledgments 

Acknowledgments: We thank the anonymous referee for the helpful
comments. We acknowledge the use of public data from the \xmm\ data
archive and the 2XMM Serendipitous Source Catalog, constructed by the
XMM-Newton Survey Science Center on behalf of ESA.

\clearpage
\setlength{\tabcolsep}{0.02in}
\begin{deluxetable}{r|c|c|cc|cccc|c|cc}
\tablecaption{Spectral fits of candidate compact objects\label{tbl:spfit}}
\tablewidth{0pt}
\tablehead{Src& Obs & $N_{\rm H}$ &
  $\Gamma_{\rm PL}$ &$N_{\rm PL}$ & 
  $E_{\rm Ga}$ & $\sigma_{\rm Ga}$ & $N_{\rm Ga}$ & EW &
   $\chi^2_\nu(\nu)$ & $F_{\rm abs}$ & $F_{\rm unabs}$\\
                &  & (10$^{22}$ cm$^{-2}$)&
         & ($10^{-5}$)&
  (keV) & (keV) & ($10^{-5}$) & (keV) &
    & \multicolumn{2}{c}{(10$^{-13}$ erg s$^{-1}$ cm$^{-2}$)}\\
 (1)&(2)&(3)&(4)&(5)&(6)&(7)&(8)&(9)&(10)&(11)&(12)
}
\startdata
\multirow{2}{*}{1}&1 & \multirow{2}{*}{$ 1.0^{+ 0.5}_{-0.4}$}  & $  1.0^{+  0.3}_{ -0.3}$  & $ 1.1^{+0.8}_{ -0.4}$ &\nodata&\nodata&\nodata&\nodata&$0.93( 25)$ & $  1.5^{+  0.3}_{ -0.2}$  & $  1.8^{+  0.3}_{ -0.3}$ \\
&2+3 &  & $  1.2^{+  0.4}_{ -0.4}$  & $ 1.1^{+0.9}_{ -0.4}$ &\nodata&\nodata&\nodata&\nodata&$1.15( 21)$ & $  1.0^{+  0.2}_{ -0.2}$  & $  1.3^{+  0.2}_{ -0.2}$ \\
\hline
\multirow{4}{*}{2}&1 & \multirow{2}{*}{$ 5.3^{+ 1.0}_{-0.9}$}  & $  0.3^{+  0.2}_{ -0.2}$  & $ 1.8^{+0.9}_{ -0.6}$ &\nodata&\nodata&\nodata&\nodata&$1.65( 93)$ & $  6.4^{+  0.5}_{ -0.5}$  & $  8.2^{+  0.6}_{ -0.6}$ \\
&2+3 & & $  0.4^{+  0.2}_{ -0.2}$  & $ 1.9^{+ 0.9}_{ -0.6}$ &\nodata&\nodata&\nodata&\nodata&$1.42( 83)$ & $  5.3^{+  0.5}_{ -0.4}$  & $  6.9^{+  0.6}_{ -0.5}$ \\
\cline{2-12}
&1 & \multirow{2}{*}{$ 5.8^{+ 1.1}_{-0.9}$}  & $  0.6^{+  0.3}_{ -0.2}$  & $ 2.7^{+ 1.6}_{ -1.0}$  & \multirow{2}{*}{$ 6.6^{+ 0.1}_{-0.1}$}  & \multirow{2}{*}{$ 0.3^{+ 0.1}_{-0.1}$}  & $ 0.8^{+ 0.2}_{-0.2}$  & $1.0^{+0.3}_{-0.3}$ &$1.02( 90)$ & $  6.2^{+  0.5}_{ -0.4}$  & $  8.4^{+  0.8}_{ -0.7}$ \\
&2+3 &  & $  0.8^{+  0.3}_{ -0.3}$  & $ 2.9^{+ 1.9}_{-1.1}$  & & & $ 0.5^{+ 0.2}_{-0.2}$  & $0.8^{+0.4}_{-0.3}$ &$1.11( 80)$ & $  5.0^{+  0.4}_{ -0.4}$  & $  7.1^{+  0.7}_{ -0.6}$ \\
\hline
\multirow{2}{*}{27}&1 & \multirow{2}{*}{$ 0.7^{+ 0.3}_{-0.2}$} & $  1.0^{+  0.3}_{ -0.3}$  & $ 1.1^{+0.6}_{ -0.4}$ &\nodata&\nodata&\nodata&\nodata & $0.55( 21)$ & $  1.4^{+  0.2}_{ -0.2}$  & $  1.7^{+  0.2}_{ -0.2}$ \\
&2+3 &  & $  1.1^{+  0.3}_{ -0.3}$  & $ 1.1^{+0.5}_{ -0.3}$ &\nodata&\nodata&\nodata&\nodata & $0.71( 24)$ & $  1.1^{+  0.2}_{ -0.2}$  & $  1.4^{+  0.2}_{ -0.2}$ \\
\hline
\multirow{2}{*}{30}&1 & \multirow{2}{*}{$ 4.8^{+ 3.4}_{-2.2}$}  & $  0.3^{+  0.7}_{ -0.6}$  & $  0.5^{+ 1.3}_{ -0.3}$ &\nodata&\nodata&\nodata&\nodata&$0.98( 13)$ & $  1.8^{+  0.4}_{ -0.4}$  & $  2.3^{+  0.5}_{ -0.4}$ \\
&2+3 &  & $  0.1^{+  0.6}_{ -0.5}$  & $  0.4^{+ 0.7}_{ -0.2}$ &\nodata&\nodata&\nodata&\nodata&$0.89( 17)$ & $  2.1^{+  0.4}_{ -0.4}$  & $  2.5^{+  0.5}_{ -0.4}$ \\
\hline
\multirow{4}{*}{34}&1 & \multirow{2}{*}{$ 3.7^{+ 1.5}_{-1.2}$}  & $  0.4^{+  0.4}_{ -0.4}$  & $ 1.1^{+ 1.2}_{ -0.5}$ &\nodata&\nodata&\nodata&\nodata&$1.50( 33)$ & $  3.4^{+  0.5}_{ -0.5}$  & $  4.2^{+  0.6}_{ -0.6}$ \\
&2+3 &  & $  0.2^{+  0.4}_{ -0.4}$  & $ 1.1^{+ 1.1}_{ -0.5}$ &\nodata&\nodata&\nodata&\nodata&$1.07( 36)$ & $  4.9^{+  0.7}_{ -0.7}$  & $  5.9^{+  0.8}_{ -0.8}$ \\
\cline{2-12}
&1 & \multirow{2}{*}{$ 5.0^{+ 2.0}_{-1.6}$}  & $  1.1^{+  0.6}_{ -0.5}$  & $ 3.0^{+ 5.5}_{-1.8}$  & \multirow{2}{*}{$ 6.7^{+ 0.1}_{-0.1}$}  & \multirow{2}{*}{$ 0.3^{+ 0.1}_{-0.1}$}  & $  0.6^{+  0.3}_{ -0.3}$  & $1.7^{+1.2}_{-0.8}$ &$1.04( 30)$ & $  3.3^{+  0.5}_{ -0.5}$  & $  4.8^{+  1.9}_{ -0.8}$ \\
&2+3 & & $  0.7^{+  0.6}_{ -0.5}$  & $ 2.3^{+ 3.6}_{-1.3}$  & & & $ 0.6^{+ 0.4}_{ -0.3}$  & $1.0^{+0.8}_{-0.5}$ &$0.79( 33)$ & $  4.5^{+  0.7}_{ -0.7}$  & $  6.2^{+  1.3}_{ -0.9}$ \\
\hline
\multirow{4}{*}{36}&1 & \multirow{2}{*}{$ 5.0^{+ 1.1}_{-0.9}$}  & $  0.5^{+  0.3}_{ -0.2}$  & $ 1.8^{+ 1.1}_{ -0.7}$ &\nodata&\nodata&\nodata&\nodata&$1.00( 63)$ & $  4.9^{+  0.5}_{ -0.5}$  & $  6.5^{+  0.6}_{ -0.6}$ \\
&2+3 & & $  0.8^{+  0.3}_{ -0.3}$  & $ 3.2^{+ 1.9}_{-1.1}$ &\nodata&\nodata&\nodata&\nodata&$1.36( 73)$ & $  4.9^{+  0.5}_{ -0.5}$  & $  7.0^{+  0.8}_{ -0.7}$ \\
\cline{2-12}
&1 & \multirow{2}{*}{$ 5.5^{+ 1.2}_{-1.0}$}  & $  0.7^{+  0.3}_{ -0.3}$  & $ 2.6^{+ 1.8}_{ -1.0}$  & \multirow{2}{*}{$ 6.7^{+ 0.1}_{-0.1}$}  & \multirow{2}{*}{$ 0.2^{+ 0.1}_{-0.1}$}  & $  0.3^{+ 0.2}_{ -0.2}$  & $0.4^{+0.3}_{-0.2}$ &$0.91( 60)$ & $  4.8^{+  0.5}_{ -0.5}$  & $  6.7^{+  0.8}_{ -0.7}$ \\
&2+3 & & $  1.1^{+  0.4}_{ -0.3}$  & $ 5.3^{+ 4.2}_{-2.2}$  & & & $  0.6^{+ 0.2}_{ -0.2}$  & $0.9^{+0.5}_{-0.4}$ &$1.06( 70)$ & $  4.8^{+  0.5}_{ -0.5}$  & $  7.6^{+  1.4}_{ -0.9}$
\enddata 
\tablecomments{Columns: (1) our source numbering; (2) the observation; (3) the absorption column density; (4)-(5) the PL photon index and normalization, respectively; (6)-(9) the centroid energy, the FWHM, the normalization, and the equivalent width of the Gaussian emission line, respectively; (10) the reduced $\chi^2$ ($\chi^2_\nu$) and the degrees of freedom ($\nu$); (11)-(12) the 0.3--10.0 keV absorbed and unabsorbed fluxes, respectively. All the fits were made to the 0.3--10.0 keV spectra. Parameters with values shared by both Obs 1 and Obs 2--3 are tied to be the same between these observations. }
\end{deluxetable}

\clearpage
{
\tabletypesize{\tiny}
\setlength{\tabcolsep}{0.005in}
\begin{deluxetable}{rlccccccccccccccccccccccccccccccccc}
\addtolength{\textwidth}{0.35cm}

\rotate
\centering
\tablecaption{General Source Properties from LWB12 \label{tbl:dercat}}
\tablewidth{0pt}
\tablehead{ & & & & & & & & & & & & & & & \multicolumn{5}{c}{Obs 1} & & \multicolumn{5}{c}{Obs 2}& & \multicolumn{5}{c}{Obs 3}\\
\cline{16-20}\cline{22-26}\cline{28-32}
\colhead{Src} & \colhead{SRCID} & \colhead{IAUNAME} &\colhead{RA} &\colhead{decl} &\colhead{$\sigma_{\rm p}$} & \colhead{Dxp} & \colhead{Dx$\gamma$} & \colhead{Dxo} & \colhead{$R2$} & \colhead{Rxo} & \colhead{Dxir} & \colhead{$K_{\rm s}$} &\colhead{Rxir}  & & S$/$N & $N_{\rm H}$ & $\Gamma_{\rm PL}$ & $F_{\rm PL}$ & $\chi^2_\nu$(dof) & &  S$/$N & $N_{\rm H}$ & $\Gamma_{\rm PL}$ & $F_{\rm PL}$ & $\chi^2_\nu$(dof) & & S$/$N & $N_{\rm H}$ & $\Gamma_{\rm PL}$ & $F_{\rm PL}$ & $\chi^2_\nu$(dof)\\
(1) &(2) &(3) &(4) &(5) &(6) &(7) &(8) &(9) &(10) &(11) &(12) &(13) &(14) & & (15)&(16) &(17) &(18)& (19)& &(20) &(21) &(22) & (23) &(24)  &&(25) &(26) &(27) &(28) & (29)
}
\startdata
1 & 254026  & J180442.0-214221  & 18:04:42.06  & -21:42:21.7  & 0.4  & 8.8  & 2.3  & \nodata & \nodata & 1.1  & \nodata & \nodata & 0.4  & &21.4  & $1.1^{+0.6}_{-0.4}$  & $0.8^{+0.4}_{-0.4}$  & $24^{+6}_{-5}$ &$0.7(10)$  & &11.4  & $1.2^{+1.4}_{-0.7}$  & $0.9^{+0.9}_{-0.6}$  & $16^{+7}_{-6}$ &$0.5(10)$  & &13.1  & $1.0^{+1.1}_{-0.5}$  & $1.2^{+0.8}_{-0.5}$  & $15^{+6}_{-5}$ &$0.7(10)$  \\
2 & 253961  & J180432.5-214009  & 18:04:32.54  & -21:40:09.2  & 0.3  & 6.1  & 2.0  & \nodata & \nodata & 1.7  & \nodata & \nodata & 1.0  & &41.5  & $5.3^{+3.8}_{-1.8}$  & $0.0^{+0.6}_{-0.5}$  & $125^{+32}_{-26}$ &$0.8(10)$  & &23.4  & $8.2^{+1.8}_{-4.0}$  & $0.5^{+0.6}_{-0.8}$  & $80^{+25}_{-16}$ &$0.4(10)$  & &27.9  & $3.8^{+4.2}_{-1.7}$  & $-0.3^{+0.7}_{-0.5}$  & $114^{+30}_{-24}$ &$0.3(10)$  \\
27 & 253711  & J180351.4-213708  & 18:03:51.45  & -21:37:08.1  & 0.4  & 4.2  & 10.7  & \nodata & \nodata & 1.0  & \nodata & \nodata & 0.3  & &24.1  & $1.6^{+0.8}_{-0.5}$  & $1.7^{+0.5}_{-0.4}$  & $28^{+6}_{-5}$ &$1.7(10)$  & &15.5  & $1.1^{+0.8}_{-0.5}$  & $1.2^{+0.6}_{-0.4}$  & $16^{+5}_{-4}$ &$2.0(10)$  & &15.6  & $2.5^{+1.3}_{-1.1}$  & $2.0^{+0.8}_{-0.7}$  & $11^{+3}_{-3}$ &$0.7(10)$  \\
30 & 253692  & J180347.9-214925  & 18:03:47.90  & -21:49:25.2  & 0.5  & 10.8  & 12.7  & \nodata & \nodata & 1.2  & \nodata & \nodata & 0.5  & &14.6  & $4.0^{+6.0}_{-3.0}$  & $-0.4^{+1.2}_{-0.8}$  & $44^{+17}_{-14}$ &$0.4(6)$  & &10.3  & $10.0^{+0.0}_{-8.8}$  & $-0.1^{+0.6}_{-1.5}$  & $42^{+23}_{-12}$ &$0.4(6)$  & &10.8  & $10.0^{+0.0}_{-8.3}$  & $0.9^{+0.6}_{-1.6}$  & $29^{+17}_{-8}$ &$0.5(6)$  \\
34 & 253618  & J180330.3-212825  & 18:03:30.37  & -21:28:25.4  & 0.3  & 13.9  & 19.9  & \nodata & \nodata & 1.5  & \nodata & \nodata & 0.8  & &22.8  & $1.9^{+1.6}_{-1.0}$  & $-0.4^{+0.5}_{-0.4}$  & $77^{+22}_{-16}$ &$0.9(10)$  & &16.4  & $2.4^{+4.6}_{-1.6}$  & $-0.4^{+0.8}_{-0.6}$  & $83^{+29}_{-20}$ &$0.7(10)$  & &16.2  & $5.1^{+4.9}_{-2.9}$  & $0.0^{+1.1}_{-0.7}$  & $70^{+27}_{-21}$ &$0.4(10)$  \\
36 & 253588  & J180323.4-213128  & 18:03:23.45  & -21:31:28.1  & 0.3  & 12.8  & 19.2  & \nodata & \nodata & 1.7  & \nodata & \nodata & 1.0  & &31.0  & $3.0^{+1.9}_{-1.2}$  & $-0.3^{+0.5}_{-0.4}$  & $103^{+27}_{-19}$ &$0.9(10)$  & &22.0  & $3.3^{+2.4}_{-1.4}$  & $0.2^{+0.6}_{-0.5}$  & $85^{+24}_{-19}$ &$0.5(10)$  & &24.8  & $2.9^{+2.9}_{-1.4}$  & $-0.4^{+0.6}_{-0.4}$  & $121^{+31}_{-25}$ &$0.9(10)$  \\
\hline
3 & 254070  & J180459.1-214138  & 18:04:59.16  & -21:41:38.2  & 0.8  & 12.4  & 6.2  & 1.0 & 17.1 & -1.5  & 0.9 & 12.5 & -1.8  & &10.2  & $1.3^{+0.1}_{-0.7}$  & $10.0^{+0.0}_{-4.8}$  & $1.0^{+0.3}_{-0.2}$ &$1.3(10)$  & &4.2  & $1.1^{+0.2}_{-0.9}$  & $10.0^{+0.0}_{-6.2}$  & $0.5^{+0.2}_{-0.2}$ &$0.3(10)$  & &4.9  & $0.8^{+0.8}_{-0.7}$  & $6.7^{+3.3}_{-4.7}$  & $0.6^{+0.8}_{-0.2}$ &$0.2(10)$  \\
4 & 254057  & J180453.3-214257  & 18:04:53.37  & -21:42:57.4  & 0.5  & 11.4  & 4.9  & 0.7 & 15.3 & -1.8  & 1.1 & 11.5 & -1.7  & &16.3  & $0.3^{+0.1}_{-0.1}$  & $2.6^{+0.6}_{-0.5}$  & $4.0^{+1.3}_{-1.0}$ &$0.9(10)$  & & \nodata  & \nodata  & \nodata  & \nodata  & \nodata  & &11.0  & $0.4^{+0.3}_{-0.2}$  & $3.0^{+1.3}_{-0.8}$  & $2.9^{+1.5}_{-0.9}$ &$0.5(10)$  \\
5 & 254042  & J180444.2-214739  & 18:04:44.26  & -21:47:39.9  & 1.0  & 12.0  & 6.2  & \nodata & \nodata & 0.7  & 1.2 & 12.7 & -1.1  & & \nodata  & \nodata  & \nodata  & \nodata  & \nodata  & & \nodata  & \nodata  & \nodata  & \nodata  & \nodata  & &9.1  & $2.5^{+4.5}_{-1.4}$  & $2.8^{+4.4}_{-1.3}$  & $5.2^{+4.1}_{-2.3}$ &$0.6(10)$  \\
6 & 254028  & J180442.1-214255  & 18:04:42.16  & -21:42:55.7  & 1.4  & 9.0  & 2.4  & 3.7 & 12.2 & -3.4  & 3.8 & 7.7 & -3.7  & & \nodata  & \nodata  & \nodata  & \nodata  & \nodata  & & \nodata  & \nodata  & \nodata  & \nodata  & \nodata  & &6.4  & $1.3^{+7.4}_{-0.8}$  & $2.9^{+7.1}_{-1.3}$  & $1.6^{+1.4}_{-0.8}$ &$0.3(10)$  \\
7 & 254015  & J180440.8-214102  & 18:04:40.88  & -21:41:02.5  & 1.3  & 8.2  & 2.2  & 3.1 & 20.2 & 0.0  & 2.9 & 13.5 & -1.1  & & \nodata  & \nodata  & \nodata  & \nodata  & \nodata  & & \nodata  & \nodata  & \nodata  & \nodata  & \nodata  & &6.5  & $4.0^{+6.0}_{-2.7}$  & $3.2^{+5.9}_{-1.9}$  & $2.6^{+2.7}_{-1.4}$ &$0.3(10)$  \\
8 & 254000  & J180438.3-214517  & 18:04:38.33  & -21:45:17.7  & 0.7  & 9.4  & 3.5  & 1.5 & 12.1 & -3.8  & 1.4 & 7.5 & -4.1  & &9.4  & $1.0^{+0.8}_{-0.4}$  & $3.3^{+1.5}_{-1.0}$  & $1.5^{+0.8}_{-0.5}$ &$0.8(10)$  & &5.6  & $2.3^{+1.1}_{-1.1}$  & $9.6^{+0.4}_{-4.4}$  & $0.7^{+0.3}_{-0.2}$ &$0.3(10)$  & &7.0  & $1.1^{+1.1}_{-0.5}$  & $4.9^{+5.1}_{-1.8}$  & $0.8^{+0.4}_{-0.3}$ &$1.1(10)$  \\
9 & 253991  & J180436.3-214306  & 18:04:36.33  & -21:43:06.5  & 1.1  & 7.8  & 1.3  & \nodata & \nodata & 0.6  & \nodata & \nodata & -0.1  & & \nodata  & \nodata  & \nodata  & \nodata  & \nodata  & &7.8  & $0.5^{+1.3}_{-0.3}$  & $0.7^{+0.9}_{-0.6}$  & $7.5^{+4.1}_{-3.1}$ &$0.7(10)$  & & \nodata  & \nodata  & \nodata  & \nodata  & \nodata  \\
10 & 253986  & J180435.9-214502  & 18:04:35.92  & -21:45:02.6  & 1.0  & 8.8  & 3.0  & 0.2 & 17.2 & -1.9  & 0.4 & 13.4 & -1.8  & &5.4  & $0.2^{+1.2}_{-0.2}$  & $2.6^{+7.4}_{-1.2}$  & $0.6^{+0.6}_{-0.3}$ &$0.8(10)$  & &4.4  & $1.5^{+0.4}_{-1.1}$  & $10.0^{+0.0}_{-6.5}$  & $0.3^{+0.2}_{-0.1}$ &$0.3(10)$  & &4.2  & $1.1^{+0.3}_{-1.1}$  & $10.0^{+0.0}_{-8.3}$  & $0.2^{+0.5}_{-0.1}$ &$0.7(10)$  \\
11 & 253978  & J180434.7-213915  & 18:04:34.79  & -21:39:15.7  & 0.8  & 6.6  & 2.9  & 2.0 & 19.4 & -0.2  & 2.1 & 7.7 & -3.3  & &7.7  & $5.6^{+4.4}_{-4.0}$  & $5.8^{+4.2}_{-3.3}$  & $1.3^{+0.8}_{-0.5}$ &$0.8(10)$  & &6.2  & $1.1^{+2.6}_{-0.9}$  & $1.1^{+1.9}_{-1.0}$  & $4.6^{+3.6}_{-2.6}$ &$0.6(10)$  & &5.9  & $4.4^{+5.6}_{-3.2}$  & $4.7^{+5.3}_{-2.8}$  & $1.3^{+1.1}_{-0.6}$ &$0.5(10)$  \\
12 & 253954  & J180432.0-213705  & 18:04:32.00  & -21:37:05.7  & 0.6  & 6.4  & 5.0  & \nodata & \nodata & -0.4  & 2.0 & 9.2 & -3.5  & &8.4  & $1.4^{+0.1}_{-0.4}$  & $10.0^{+0.0}_{-2.7}$  & $0.5^{+0.1}_{-0.1}$ &$1.2(10)$  & &6.9  & $1.4^{+0.2}_{-0.9}$  & $10.0^{+0.0}_{-5.7}$  & $0.5^{+0.2}_{-0.2}$ &$1.1(10)$  & &7.9  & $0.5^{+1.2}_{-0.2}$  & $3.8^{+6.2}_{-1.3}$  & $1.2^{+0.6}_{-0.5}$ &$1.1(10)$  \\
13 & 253929  & J180428.4-213349  & 18:04:28.48  & -21:33:49.8  & 0.7  & 7.7  & 8.3  & 1.5 & 14.0 & -3.3  & 1.3 & 12.5 & -2.3  & &8.3  & $1.5^{+0.2}_{-0.7}$  & $10.0^{+0.0}_{-3.8}$  & $0.4^{+0.1}_{-0.1}$ &$0.2(10)$  & &4.6  & $1.6^{+0.3}_{-0.7}$  & $10.0^{+0.0}_{-4.0}$  & $0.4^{+0.2}_{-0.2}$ &$0.6(10)$  & &4.2  & $1.3^{+0.2}_{-0.7}$  & $10.0^{+0.0}_{-4.2}$  & $0.3^{+0.1}_{-0.1}$ &$0.5(10)$  \\
14 & 253916  & J180426.6-213732  & 18:04:26.65  & -21:37:32.3  & 0.6  & 5.1  & 4.8  & 0.2 & 12.0 & -4.0  & 0.5 & 10.6 & -3.0  & &8.3  & $1.2^{+0.1}_{-0.4}$  & $10.0^{+0.0}_{-2.8}$  & $0.4^{+0.1}_{-0.1}$ &$2.3(10)$  & &7.0  & $1.1^{+0.1}_{-0.5}$  & $10.0^{+0.0}_{-3.3}$  & $0.6^{+0.2}_{-0.2}$ &$1.2(10)$  & &6.8  & $1.3^{+0.2}_{-0.5}$  & $10.0^{+0.0}_{-3.3}$  & $0.4^{+0.2}_{-0.1}$ &$0.6(10)$  \\
15 & 253895  & J180422.9-214155  & 18:04:22.97  & -21:41:55.6  & 1.1  & 4.5  & 2.2  & \nodata & \nodata & 0.2  & \nodata & \nodata & -0.5  & &7.6  & $2.5^{+3.7}_{-1.8}$  & $1.4^{+1.4}_{-1.0}$  & $3.2^{+1.8}_{-1.3}$ &$0.2(10)$  & & \nodata  & \nodata  & \nodata  & \nodata  & \nodata  & & \nodata  & \nodata  & \nodata  & \nodata  & \nodata  \\
16 & 253885  & J180421.5-214233  & 18:04:21.52  & -21:42:33.3  & 0.4  & 4.6  & 2.6  & 0.5 & 16.4 & -1.1  & 0.5 & 11.9 & -1.4  & &28.2  & $0.6^{+0.1}_{-0.1}$  & $2.2^{+0.3}_{-0.3}$  & $8.1^{+1.6}_{-1.4}$ &$1.4(10)$  & &10.3  & $0.4^{+0.3}_{-0.2}$  & $2.1^{+0.7}_{-0.6}$  & $2.3^{+1.2}_{-0.7}$ &$1.4(10)$  & &10.6  & $0.9^{+0.5}_{-0.3}$  & $2.5^{+1.3}_{-0.8}$  & $2.2^{+1.4}_{-0.9}$ &$1.7(10)$  \\
17 & 253884  & J180421.4-212956  & 18:04:21.49  & -21:29:56.1  & 1.1  & 10.2  & 12.4  & 1.7 & 19.3 & -0.5  & 2.0 & 13.0 & -1.5  & &10.9  & $0.9^{+1.0}_{-0.5}$  & $5.7^{+4.3}_{-2.8}$  & $0.8^{+0.5}_{-0.3}$ &$6.1(10)$  & & \nodata  & \nodata  & \nodata  & \nodata  & \nodata  & & \nodata  & \nodata  & \nodata  & \nodata  & \nodata  \\
18 & 253874  & J180419.5-213604  & 18:04:19.60  & -21:36:04.2  & 0.7  & 4.6  & 6.7  & 0.3 & 19.2 & -0.9  & 1.0 & 12.0 & -2.2  & &11.5  & $0.7^{+0.3}_{-0.2}$  & $3.3^{+1.0}_{-0.7}$  & $1.2^{+0.5}_{-0.4}$ &$0.4(10)$  & & \nodata  & \nodata  & \nodata  & \nodata  & \nodata  & & \nodata  & \nodata  & \nodata  & \nodata  & \nodata  \\
19 & 253850  & J180416.3-212949  & 18:04:16.33  & -21:29:49.5  & 0.6  & 10.0  & 12.9  & 0.8 & 14.9 & -2.7  & 0.9 & 12.4 & -2.1  & &11.8  & $1.3^{+0.1}_{-0.2}$  & $10.0^{+0.0}_{-1.2}$  & $0.9^{+0.2}_{-0.2}$ &$1.5(10)$  & &7.4  & $1.4^{+0.2}_{-0.6}$  & $10.0^{+0.0}_{-3.2}$  & $0.7^{+0.2}_{-0.2}$ &$0.9(10)$  & &7.1  & $1.3^{+0.2}_{-0.4}$  & $10.0^{+0.0}_{-2.3}$  & $0.5^{+0.2}_{-0.1}$ &$1.4(10)$  \\
20 & 253829  & J180413.3-213151  & 18:04:13.35  & -21:31:52.0  & 1.0  & 7.9  & 11.2  & 1.3 & 15.9 & -2.2  & 1.2 & 13.4 & -1.7  & &10.3  & $0.8^{+0.8}_{-0.4}$  & $6.0^{+4.0}_{-1.8}$  & $0.8^{+0.2}_{-0.2}$ &$2.4(10)$  & & \nodata  & \nodata  & \nodata  & \nodata  & \nodata  & & \nodata  & \nodata  & \nodata  & \nodata  & \nodata  \\
21 & 253814  & J180409.9-214738  & 18:04:09.99  & -21:47:38.7  & 0.7  & 8.1  & 7.6  & \nodata & \nodata & 0.6  & \nodata & \nodata & -0.1  & & \nodata  & \nodata  & \nodata  & \nodata  & \nodata  & &5.5  & $1.2^{+5.9}_{-1.1}$  & $0.4^{+1.6}_{-1.1}$  & $11^{+8}_{-6}$ &$0.5(6)$  & &9.7  & $0.8^{+1.7}_{-0.6}$  & $0.3^{+0.8}_{-0.6}$  & $24^{+11}_{-9}$ &$0.7(6)$  \\
22 & 253797  & J180406.5-214519  & 18:04:06.52  & -21:45:19.6  & 0.5  & 5.8  & 6.8  & \nodata & \nodata & 0.1  & 0.7 & 13.1 & -1.5  & &14.2  & $0.5^{+0.3}_{-0.2}$  & $3.9^{+1.5}_{-0.8}$  & $1.4^{+0.4}_{-0.3}$ &$0.7(6)$  & &7.2  & $0.5^{+1.1}_{-0.3}$  & $4.4^{+5.6}_{-1.9}$  & $0.8^{+0.5}_{-0.3}$ &$0.9(6)$  & &9.3  & $1.5^{+0.2}_{-0.8}$  & $10.0^{+0.0}_{-4.7}$  & $0.7^{+0.2}_{-0.2}$ &$1.0(6)$  \\
23 & 253767  & J180400.7-214252  & 18:04:00.73  & -21:42:52.5  & 0.3  & 3.6  & 7.4  & 0.5 & 13.2 & -2.6  & 0.2 & 10.2 & -2.2  & &36.9  & $0.6^{+0.6}_{-0.2}$  & $6.0^{+4.0}_{-1.1}$  & $3.9^{+0.6}_{-0.7}$ &$3.8(10)$  & &27.6  & $0.4^{+0.2}_{-0.1}$  & $4.2^{+1.0}_{-0.6}$  & $4.5^{+0.8}_{-0.7}$ &$3.8(10)$  & &23.2  & $1.3^{+0.1}_{-0.9}$  & $10.0^{+0.0}_{-5.8}$  & $2.6^{+1.0}_{-0.4}$ &$2.5(10)$  \\
24 & 253756  & J180358.9-214248  & 18:03:58.95  & -21:42:48.2  & 1.3  & 3.7  & 7.8  & 2.0 & 15.7 & -2.4  & 2.5 & 13.6 & -1.7  & &8.1  & $0.2^{+0.3}_{-0.2}$  & $2.3^{+1.9}_{-0.8}$  & $0.7^{+0.5}_{-0.4}$ &$2.4(10)$  & & \nodata  & \nodata  & \nodata  & \nodata  & \nodata  & & \nodata  & \nodata  & \nodata  & \nodata  & \nodata  \\
25 & 253754  & J180358.7-212725  & 18:03:58.71  & -21:27:25.1  & 0.5  & 12.3  & 16.7  & 0.7 & 16.8 & -1.0  & 0.7 & 12.8 & -1.0  & &25.0  & $0.3^{+0.1}_{-0.1}$  & $2.7^{+0.5}_{-0.4}$  & $6.0^{+1.4}_{-1.1}$ &$2.3(10)$  & & \nodata  & \nodata  & \nodata  & \nodata  & \nodata  & & \nodata  & \nodata  & \nodata  & \nodata  & \nodata  \\
26 & 253719  & J180353.0-214950  & 18:03:53.08  & -21:49:50.9  & 0.8  & 10.8  & 12.0  & 0.2 & 15.3 & -2.2  & 0.2 & 11.9 & -2.0  & &7.4  & $1.3^{+0.2}_{-0.9}$  & $10.0^{+0.0}_{-5.9}$  & $0.5^{+0.1}_{-0.1}$ &$0.8(6)$  & &9.2  & $0.3^{+0.5}_{-0.2}$  & $3.3^{+3.2}_{-1.0}$  & $1.9^{+1.0}_{-0.7}$ &$0.9(6)$  & &5.8  & $1.4^{+0.2}_{-0.8}$  & $10.0^{+0.0}_{-5.2}$  & $0.5^{+0.2}_{-0.1}$ &$2.2(6)$  \\
28 & 253700  & J180349.4-214136  & 18:03:49.40  & -21:41:36.1  & 0.5  & 4.4  & 10.0  & 0.6 & 14.7 & -2.8  & 0.8 & 13.2 & -1.8  & &14.5  & $1.5^{+0.1}_{-0.7}$  & $10.0^{+0.0}_{-4.2}$  & $0.8^{+0.2}_{-0.2}$ &$1.0(10)$  & &9.2  & $1.3^{+0.1}_{-0.6}$  & $10.0^{+0.0}_{-3.9}$  & $0.6^{+0.2}_{-0.1}$ &$1.0(10)$  & &9.5  & $0.6^{+0.9}_{-0.3}$  & $4.9^{+5.1}_{-1.5}$  & $0.8^{+0.3}_{-0.2}$ &$0.7(10)$  \\
29 & 253695  & J180348.8-213034  & 18:03:48.84  & -21:30:34.2  & 0.5  & 9.9  & 15.4  & 0.4 & 12.5 & -3.7  & 1.1 & 11.7 & -2.4  & &10.9  & $1.4^{+0.1}_{-0.3}$  & $10.0^{+0.0}_{-1.8}$  & $0.6^{+0.1}_{-0.1}$ &$1.9(10)$  & &8.0  & $1.3^{+0.1}_{-0.5}$  & $10.0^{+0.0}_{-3.0}$  & $0.7^{+0.2}_{-0.2}$ &$1.0(10)$  & &8.9  & $1.4^{+0.1}_{-0.3}$  & $10.0^{+0.0}_{-2.0}$  & $0.8^{+0.2}_{-0.2}$ &$2.3(10)$  \\
31 & 253677  & J180345.3-213039  & 18:03:45.36  & -21:30:39.2  & 0.5  & 10.1  & 15.8  & 0.3 & 13.5 & -3.3  & 0.4 & 12.2 & -2.2  & &11.2  & $0.6^{+1.0}_{-0.3}$  & $4.8^{+5.2}_{-1.6}$  & $1.0^{+0.4}_{-0.3}$ &$1.5(10)$  & &12.0  & $1.4^{+0.1}_{-1.0}$  & $10.0^{+0.0}_{-6.3}$  & $1.3^{+0.6}_{-0.2}$ &$0.9(10)$  & &9.7  & $1.7^{+0.2}_{-0.7}$  & $10.0^{+0.0}_{-4.2}$  & $0.8^{+0.2}_{-0.2}$ &$1.2(10)$  \\
32 & 253658  & J180341.7-214032  & 18:03:41.71  & -21:40:32.6  & 0.5  & 5.8  & 11.9  & 1.0 & 12.2 & -3.9  & 3.2 & 11.6 & -2.6  & &12.8  & $1.3^{+0.1}_{-0.2}$  & $10.0^{+0.0}_{-1.4}$  & $0.6^{+0.1}_{-0.1}$ &$2.9(10)$  & &9.8  & $1.2^{+0.1}_{-0.4}$  & $10.0^{+0.0}_{-2.5}$  & $0.6^{+0.1}_{-0.1}$ &$2.3(10)$  & &8.9  & $1.2^{+0.1}_{-0.5}$  & $10.0^{+0.0}_{-3.0}$  & $0.5^{+0.2}_{-0.1}$ &$1.0(10)$  \\
33 & 253635  & J180336.9-214059  & 18:03:37.00  & -21:40:59.9  & 0.6  & 7.0  & 12.9  & 1.3 & 17.2 & -1.8  & 0.4 & 13.0 & -1.9  & &7.5  & $1.4^{+0.2}_{-0.9}$  & $10.0^{+0.0}_{-5.1}$  & $0.2^{+0.1}_{-0.1}$ &$1.2(10)$  & &8.4  & $0.8^{+0.7}_{-0.4}$  & $6.6^{+3.4}_{-2.5}$  & $0.7^{+0.2}_{-0.2}$ &$0.5(10)$  & &6.8  & $0.7^{+0.8}_{-0.4}$  & $6.5^{+3.5}_{-2.7}$  & $0.4^{+0.2}_{-0.1}$ &$0.3(10)$  \\
35 & 253594  & J180323.9-214110  & 18:03:23.91  & -21:41:10.2  & 0.5  & 10.0  & 15.9  & 0.5 & 15.6 & -2.4  & 0.5 & 13.2 & -1.8  & &14.0  & $0.2^{+0.2}_{-0.1}$  & $2.5^{+1.0}_{-0.6}$  & $1.9^{+1.0}_{-0.6}$ &$2.2(10)$  & &6.3  & $1.5^{+0.3}_{-1.1}$  & $10.0^{+0.0}_{-6.1}$  & $0.6^{+0.3}_{-0.2}$ &$0.5(10)$  & &10.4  & $0.3^{+0.3}_{-0.2}$  & $2.6^{+1.6}_{-0.8}$  & $1.7^{+1.0}_{-0.7}$ &$3.2(10)$  \\
37 & 253574  & J180321.5-213243  & 18:03:21.53  & -21:32:44.0  & 0.4  & 12.5  & 19.0  & 1.2 & 11.9 & -3.4  & 1.1 & 11.0 & -2.2  & &16.7  & $0.6^{+0.5}_{-0.2}$  & $4.9^{+2.8}_{-1.1}$  & $1.8^{+0.5}_{-0.4}$ &$1.3(10)$  & &13.8  & $1.2^{+0.7}_{-0.4}$  & $7.1^{+2.9}_{-1.9}$  & $1.8^{+0.5}_{-0.4}$ &$1.0(10)$  & &12.0  & $0.7^{+0.9}_{-0.3}$  & $4.9^{+5.1}_{-1.5}$  & $1.7^{+0.6}_{-0.5}$ &$1.0(10)$  \\
38 & 253543  & J180316.5-213645  & 18:03:16.52  & -21:36:45.9  & 0.7  & 11.9  & 18.4  & 0.5 & 14.6 & -2.8  & 0.6 & 10.6 & -2.9  & &8.3  & $1.3^{+0.1}_{-0.3}$  & $10.0^{+0.0}_{-1.6}$  & $0.3^{+0.1}_{-0.1}$ &$3.3(10)$  & &8.1  & $1.5^{+0.2}_{-0.4}$  & $10.0^{+0.0}_{-2.1}$  & $0.9^{+0.3}_{-0.2}$ &$0.7(10)$  & & \nodata  & \nodata  & \nodata  & \nodata  & \nodata 
\enddata 
\tablecomments{Columns: (1) Our source numbering; (2) 2XMMi-DR3 unique source identifier; (3) 2XMMi-DR3 source designation, all with prefix ``2XMMi''; (4)-(6) Mean source right ascension, mean source declination (ICRS) and 1$\sigma$ positional error (they are SC\_RA, SC\_DEC and SC\_POSERR in the 2XMMi-DR3 catalog, respectively); (7) off-axis angle (arcmin) (8) separation from \object{HESS J1804-216} (arcmin); (9) X-ray-optical separation (arcsec); (10) $R2$-band magnitude; (11) X-ray-to-optical flux ratio logarithm $\log(F_{\rm X}/F_{\rm O})$, where $F_{\rm X}$ is the maximum 0.2-12.0 keV flux among the three observations and $F_{\rm O}$ is the optical flux defined as $\log(F_{\rm O})=-R2/2.5-5.37$ (when no optical counterpart, we assumed $R2=21$; see LWB12); (12) X-ray-IR separation (arcsec); (13) $K_{\rm s}$-band magnitude; (14) X-ray-to-IR flux ratio logarithm $\log(F_{\rm X}/F_{\rm IR})$, where $F_{\rm IR}$ is the IR flux defined as $\log(F_{\rm IR})=-K_{\rm s}/2.5-6.95$ (when no IR counterpart, we assumed $K_{\rm s}=15.3$, the 3-$\sigma$ limiting sensitivity of the $K_{\rm s}$ band in the 2MASS; see LWB12); (15) S/N defined as the ratio between EP\_8\_CTS and EP\_8\_CTS\_ERR from the 2XMMi-DR3 catalog for Obs 1; (16)--(19) The results of the fits with an absorbed PL ($N_{\rm H}$, in units of $10^{22}$ cm$^{-2}$, was constrained to be $\le 10^{23}$ cm$^{-2}$ and $\Gamma_{\rm PL}$ to be $\le$10 in the fits, and $F_{\rm PL}$ is the 0.2--12.0 keV absorbed flux (10$^{-14}$ erg s$^{-1}$ cm$^{-2}$)); (20)--(24) and (25)--(29) similar to (16)--(19) but for Obs 2 and Obs 3, respectively. The six sources at the top are candidate compact object systems, while the other 32 are candidate normal stars. Src 32 was resolved by {\it Chandra} into two point sources separated by 6\farcs6, i.e., \object{CXO J180341.5-214034} (slightly brighter) and \object{CXO J180341.7-214027}, from the {\it Chandra} Source Catalog \citep{evprgl2010}. They match well in position (within 1$\arcsec$) with two similarly bright IR sources, respectively. We select the closer one to Src 32 as its counterpart, though the seperation is large (3\farcs2). In the USNO B-1.0 Catalog, these two IR sources are not deblended; thus the optical counterpart used refers to their combination.}
\end{deluxetable}
}

\begin{figure*}
\centering
\includegraphics[width=0.7\textwidth]{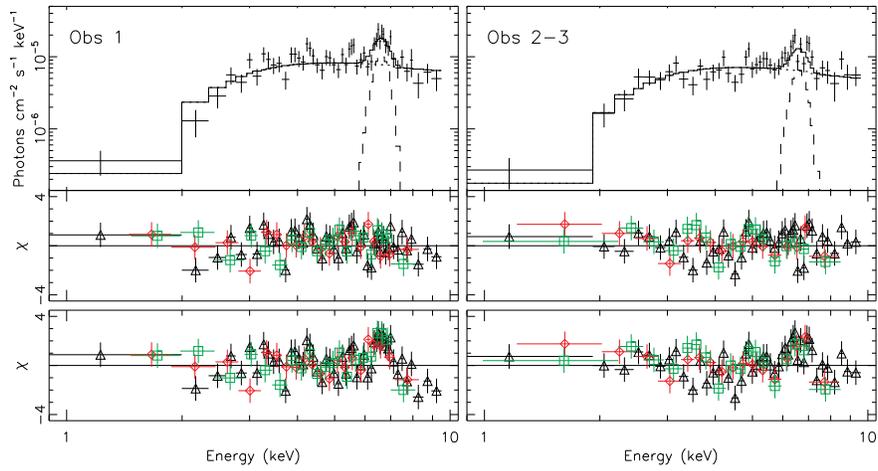}
\caption{The unfolded spectra (top panels) and the fit residuals (middle panels) using an absorbed PL plus a Gaussian Fe line model and the fit residuals using only an absorbed PL (bottom panels) for Src 2. For clarity, only the pn spectra are shown for the unfolded spectra. The dotted, dashed, and solid lines are for the PL and Fe components and the total model, respectively. The residuals are shown for all three cameras (black triangles/red diamonds/green squares for pn/MOS1/MOS2, respectively).\label{fig:s253961spfits}}
\end{figure*}

\begin{figure*} 
\centering
\includegraphics[width=6.8in]{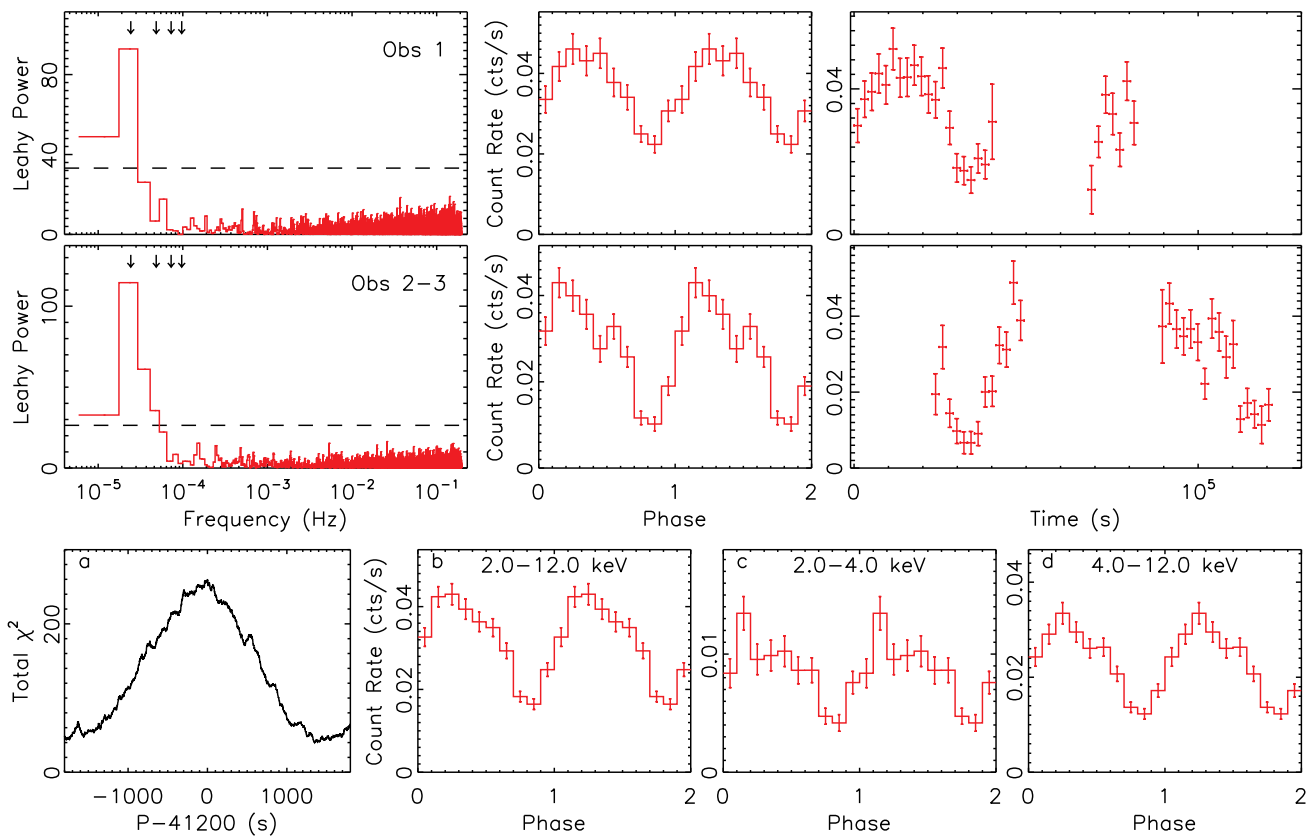}
\caption{The timing properties of Src 2, similar to
  Figure~\ref{fig:254026foldcurve}. The energy band 2.0--12.0 keV is
  used for all plots except panels (c) and (d). The period used for
  all folded light curves is 41200 s \label{fig:253961foldcurve}}
\end{figure*}

\begin{figure*} 
\centering
\includegraphics[width=7.2in]{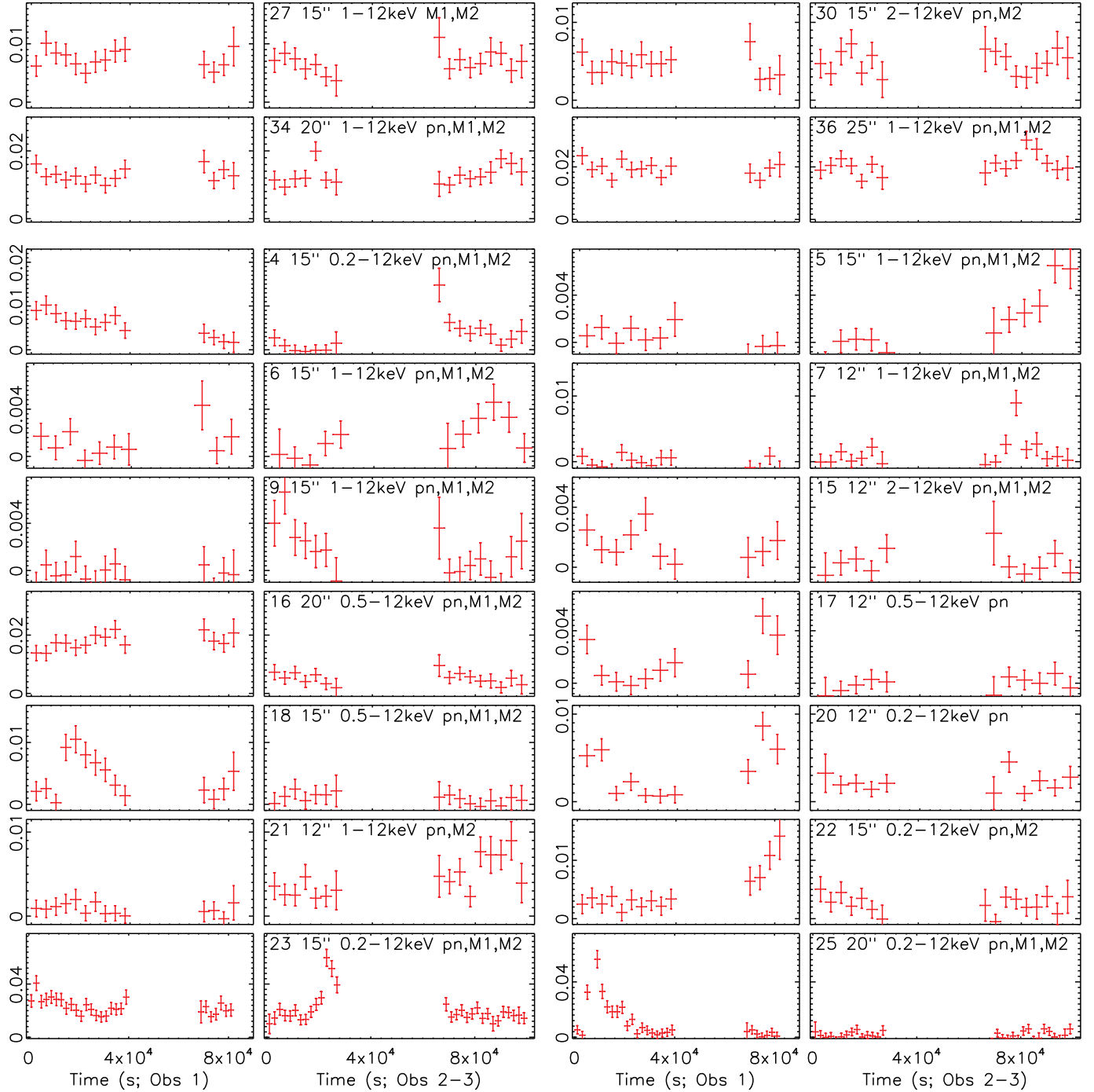}
\caption{The light curves (in units of cts s$^{-1}$) of candidate compact object systems (the top four sources), which show little short-term varibility, and those of stars with some flaring activity (others). The note in the panel containing Obs 2-3 includes the source number, the radius of the source extraction region, the energy band, and the instruments used. \label{fig:lcsum}}
\end{figure*}

\begin{figure*} 
\centering
\includegraphics[width=6.8in]{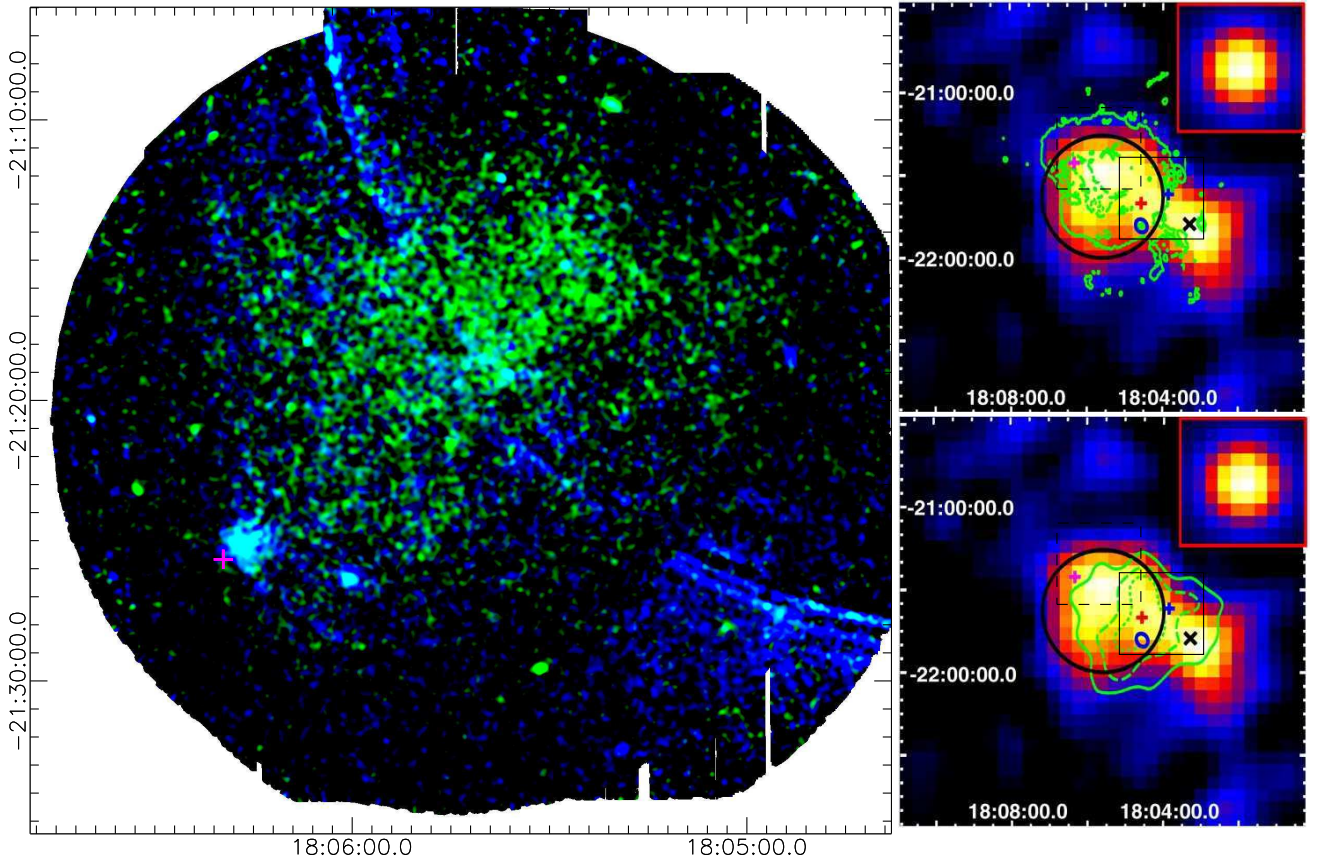}
\caption{The \xmm\ X-ray image of the field around HESS~J1804-216, similar to Figure~\ref{fig:colorimage}, but for observation 0405750201. The black dashed-line and solid-line boxes correspond to the size of the left panel of this figure and that of the left panel of Figure~\ref{fig:colorimage}, respectively. \label{fig:colorimage_ne}}
\end{figure*}

\end{document}